\documentclass[fleqn,twoside]{article}
\usepackage{espcrc2}
\newcommand{\BEQ}{\begin{equation}}
\newcommand{\EEQ}{\end{equation}}

\newcommand{\bra}[1]{\mbox{$\langle #1 \left| \right. $}}
\newcommand{\ket}[1]{\mbox{$ \left| \right.  #1 \rangle$}}

\newcommand{\VIZ}{\mbox{\em viz.\/ }}

\newcommand{\EG}{\mbox{\em e.g.\/ }}
\title{TeV String Theories, Mini Black Holes and Trans-GZK Cosmic Rays}
\author{G. Domokos\address{Department  
of Physics and Astronomy\\ The
Johns Hopkins University, Baltimore, MD}\thanks{Speaker. e-mail: {\tt Gabor.Domokos@jhu.edu}\hspace{1em} Invited talk given at XIII ISVHECRI, Pylos, Greece}
and 
S.~Kovesi-Domokos\addressmark \thanks{e-mail: {\tt skd@jhu.edu}}}
\begin{document}

\begin{abstract}
We review the proposal that trans-GZK cosmic ray interactions are caused 
by neutrino primaries. The primaries cause excitations of strings and give
rise to extensive air showers (EAS) resembling EAS induced by nuclei.
We also show that in  ``low scale'' ($M_{\ast} \approx 70$TeV) string models 
the excited string  and the mini black hole pictures are equivalent.
\end{abstract}

\maketitle

\section{Introduction}
The existence of cosmic rays transcending the Greisen-Zatsepin-Kuzmin (GZK)
cutoff appear to be a fact of  life. While there is still some controversy 
regarding the  observational data, it appears that opinions
converge toward acknowledging the existence of trans-GZK events. Clearly,
the issue will be satisfactorily clarified once future detectors, such as
OWL, EUSO, the Pierre Auger Observatory, various neutrino telescopes,
{\em etc.} will begin to take data. In the meantime, one may rely upon the
existing set of observations and seek an understanding of the phenomenon.
It is to be emphasized that  if indeed trans-GZK cosmic ray events exist,
they pose a serious challenge either to our current understanding of particle 
physics, or of astrophysics (or, perhaps, both). 
\begin{itemize}
\item From the particle physics point of view, it is to be noted that the 
well-known GZK cutoff is a {\em low energy effect}: at the GZK energy
(of a few times $10^{19}$eV), the average CMS energy in the interaction of
a high energy proton with a photon in the CMB is of the order of the mass
of the $\Delta$ resonance. This energy region has been extensively studied for
the past half century or so and it is well understood.
\item From the astrophysics point of view, starting with the famous search of
Elbert and Sommers~\cite{elbertsommers} for possible sources of trans-GZK 
cosmic rays within the the so called GZK sphere, astrophysicists have 
been looking at the sky, but, in essence, found no good candidates. 
The astrophysics problem has been somewhat aggravated by the work of Stanev
{\em et al.}~\cite{protheroe}. Those authors showed that random intergalactic
magnetic fields cause the high energy protons to undergo a Brownian motion. As
a consequence, the  distance along the line of sight of a candidate source
for emitting trans-GZK protons has to be less than  about $14$Mpc as 
opposed to $\approx 50$Mpc as previously believed on the basis of assuming
propagation along a straight line. 
\end{itemize}

There has been no dearth of theoretical schemes proposing to understand
the existence of trans-GZK CR events. In fact, currently, the number of 
theoretical papers on this subject is about an order of magnitude larger then
the number of such events observed. This is clearly a very unhealthy situation;
however, it will change once the new detectors start collecting data.

Fortunately, a substantial number of the proposed explanations is {\em dead}:
they are, by now,  excluded by accelerator based data, data on the CMB,
{\em etc.}
See the talk of S.~Kovesi-Domokos~\cite{susan} discussing this issue.

In this talk  we mainly follow a recent paper of ours~\cite{strings}, where a
more detailed discussion of the issues and a more complete list of references can be found.
\section{The Survivors}
 In essence, there are two proposals which have  not been excluded by new data or 
consistency requirements. Both of them assume that the primaries of trans-GZK 
cosmic rays are neutrinos\footnote{It is interesting to remember that neutrino
primaries for the extreme energy cosmic rays were first proposed by G.~Cocconi
at one of the early Texas Symposia (1967) and soon thereafter by
Berezinsky and Zatsepin \cite{berezin}, based on the Fermi theory
of weak interactions. Unfortunately, it appears that Cocconi's conjecture
has not  been recorded.}   
and both of them rely upon the notion that 
the characteristic scale of the ``new physics'' beyond the Standard Model
can be in the TeV range, instead of $10^{13}$ to $10^{16}$TeV~\cite{lowscale}.
\begin{itemize}
\item The excited string picture\cite{prl}
\item The mini black hole picture as applied to cosmic rays\cite{blackholes} 
\end{itemize}
Let us briefly review the main problems  any successful model of trans-GZK
cosmic rays should solve.
\begin{itemize}
\item {\em The penetration problem}. In the absence of nearby sources, one has to make 
sure that the primary
can penetrate the CMB over large distances. For neutrinos, this requirement is 
beautifully satisfied: in an interaction of a neutrino having an energy, say,
$10^{20}$eV, with a typical CMB photon,  the CMS energy is of the order of 
a few hundred MeV. Consequently, one does not even have to now about the existence
of the $W$ and $Z$ bosons.  This is {\em low energy physics}: the mean free path of the 
neutrino is comparable with the horizon size. By contrast, in an interaction
with a nucleus in the atmosphere, the CMS energy is in the range of hundreds of TeV.
Thus, if indeed the onset of the ``new physics'' is around  a  TeV to a few tens of a TeV,
one is well in the new physics regime there.
\item {\em The interaction problem.} All observed trans-GZK events are ``hadron-like'':
it appears that that the observed showers have a normal development, with an $X_{\rm max}$
(where known) as in a hadronic shower. Hence, if the primaries are neutrinos, the  new physics 
has to  have  the property that in the new physics regime, neutrinos have a large cross 
section,  about the size of a hadronic one. This is plausible: in string models, 
once the string is excited, the interactions are unified: all interactions have the 
same strength. 
\end{itemize}
The surprising fact is that the black hole and string pictures are, in essence, different
facets  of the same scheme, even though they seem to be based on very different physics.

{\em How can this happen?}
\section{Strings and Black Holes: the Equivalence}
Let us illustrate the situation on a simple example. Suppose that we want to produce a 
highly excited string in a neutrino -- quark interaction. (This is possible, since the 
excited string is unstable and it will eventually decay: there is no conflict with 
the conservation of energy-momentum.) Apart from trivial factors, the probability of
this process is given by:
\begin{equation}
P \propto \sum_{\alpha} \bra{i}{\mathcal O^{\dag}}\ket{N, \alpha}\bra{N, \alpha} {\mathcal O}\ket{i}\delta 
\left( E - NM_{\ast}\right)
\label{transitionprobability}
\end{equation}
In eq.~(\ref{transitionprobability}) $\alpha$ stands for the collection of labels necessary for the full
specification of states at level $N$: one has to sum over (most of) those, since the 
quantum numbers carried by \ket{i} and ${\mathcal O}$ do not fully specify the substate at level $N$. (We remember that,  apart from a constant -- the Regge intercept -- the mass of
string levels is given by $M^{2}\sim M_\ast^2 N$, where $N$ is a positive integer. Everywhere, $M_{\ast}$ is the characteristic
energy scale of the string model.)

We  recognize that the quantity in eq.~(\ref{transitionprobability}),
\begin{equation}
\rho_{M}= \sum_{\alpha}\ket{N, \alpha}\bra{N, \alpha} \delta \left( E - NM_{\ast} \right)
\label{microcanonical}
\end{equation}
is just the microcanonical  density matrix of the final state. 
(In statistical mechanics, a smoothing of
the delta function is necessary in order to define a level density; in the present context, 
such a smoothing is automatic if the finite width of resonances is taken into account.) 
The summation
over $\alpha$ leads to the information loss discussed by Amati~\cite{amati}characteristic of
a black hole.

Now, it is intuitively clear that a {\em large} microcanonical ensemble is like a 
canonical one: a sufficiently large system acts like its own thermal reservoir.
(In ref.~\cite{strings} we gave a formal proof of this statement, using a saddle point expansion.) 
In fact, the temperature of the large
ensemble is given by the usual formula all of us  learned in an  undergraduate course 
on statistical mechanics, \VIZ
\begin{equation}
\frac{1}{T} = \frac{\partial S}{\partial E},
\end{equation}
where $S$ (the entropy) is just the logarithm of the level density as a function of 
the energy of excitation.

Thus, for a highly excited string, we may assign a temperature to it. Also, since the
level density grows as the exponential of the energy of excitation, there is a 
limiting temperature to which the system converges in the limit of very high (``infinite'')
excitations. Thus, at very high excitations, $\mbox{strings}\cong \mbox{ black holes}$.

One might say that this is a neat theoretical construction, perhaps satisfying to the purists
who do not like two solutions to the same problem, but {\em does it have observable 
consequences}? Remarkably, the answer is ``{\em yes}''.  We now turn to explore the 
consequences of the equivalence of the string and black hole pictures.
\section{Observable Consequences}
Consider level densities of the asymptotic form:
\begin{equation}
d(E)=\exp S(E)  \sim  C \left( E/M_{\ast}\right)^{-\gamma}\exp \left(\alpha (E/M_{\ast}\right).
\label{leveldensity} 
\end{equation}

The quantities $ C, \gamma,  \alpha,  \delta $ depend on the specific model 
considered. This asymptotic form of the level density comprises all known
string models.

We now evaluate the microcanonical density matrix , eq.~(\ref{microcanonical}) by means of the saddle point method.
  
The inverse temperature turns out to be:
\begin{equation}
\beta = \frac{\partial S}{\partial E} \sim  - \frac{\gamma + 3/2 }{E} + \alpha M_{\ast}^{-1}
\label{inversetemperature}
\end{equation}
The temperature is asymptotically constant and, hence, its limit as $E\rightarrow \infty$  may be identified with the Hagedorn 
temperature. It is necessary to take Kaluza-Klein (KK) excitations into account: otherwise,
the Hagedorn temperature is not the {\em maximal}, but a {\em minimal} temperature, which is
clearly unphysical. The contribution of the KK excitations to the level density  is 
proportional to $(E/M_{\ast})^{n}$, n being the number of compactified  dimensions. The expression of 
the Hagedorn temperature 
in terms of $M_{\ast}$ is somewhat model dependent: in the calculations cited above, one has,
$T_{H}= M_{\ast}/a $, with $a$ ranging, approximately, between 2 and 4 for 
various string models. For purposes of numerical estimates, we take $a=3$.

An estimate of the single particle inclusive cross section was  given  in
ref.~\cite{strings} and we refer the reader to that article for details. 
The basis of the estimate is that in the rest frame of the excited resonance (a leptoquark 
in our case)
the distribution of the observed particle is given by a Fermi or Bose distribution,
respectively.

Here we summarize the salient features of the inclusive distribution as measured 
in the laboratory frame.
\begin{itemize}
\item The distributions are strongly peaked in the forward direction; they are 
{\em exponentially decreasing with the emission angle of the observed particle}.

In the early days of dual resonance models ({\em ca.} 1968), Gabriele Veneziano pointed 
out that the fixed angle elastic scattering amplitude was exponentially small 
away from the forward direction. We now see that this result is much more general and it is, 
in fact, independent of the tree approximation to a dual amplitude.
\item The total particle multiplicities can be estimated by creating a model for the
hadronization of quarks and gluons, for instance, as it was done in ref.~\cite{strings}.
One arrives at the conclusion that total multiplicities resemble hadron  multiplicities
created by an incoming heavy nucleus, \EG Fe. Of course, since at high string excitations 
the excited string behaves as a system in a heat bath at the Hagedorn temperature,
the relative multiplicity of any particle species is proportional to its statistical
weight. Hence, a number of prompt leptons is also created, with a smaller (by about a factor
of 3, because of the absence of color)
statistical weight. However, once one begins a realistic  simulation of $\nu$ 
induced air showers
in this picture, there may be a noticeable (and, perhaps, observable) difference in shower
development. At this meeting, Claudio Coriano presented the results of a simulation of showers 
using the semiclassical black hole picture \cite{coriano}. It is likely that an adaptation of the simulation
to the scheme presented in this talk will yield interesting and testable results.

\item In order to assess the importance of the multiplicity question, let us adopt the  
approximations made in ref.~\cite{strings} and list the hadronic multiplicities as a
function of $n$, the number of compactified dimensions. (Recall that for superstrings, $n=6$.)
The result is summarized below.
\begin{table}[h]
\begin{center}
\begin{tabular}{|c||c|c|c|}\hline
n & 6 & 7 & 8\\ \hline
$N_{had}$ & $ 1.76\times 10^{4}$ & $4.16\times 10^{3}$& $1.12\times 10^{3}$ \\ \hline
\end{tabular}
\end{center}
\label{multiplicity}
\caption{Hadron multiplicity for various numbers of extra dimensions}
\end{table}

We used a value of the characteristic string scale $M_{\ast}\approx 80$TeV, as estimated in
\cite{burgettetal}. This result is to be compared with an extrapolation of hadronic 
multiplicities assuming that there is no new physics between present day accelerator energies
and the energies of trans-GZK cosmic rays. This is relatively easy: most data listed in 
the particle data group can be fitted by means of a quadratic polynomial in $\log s$. In this 
way, one obtains a baseline: multiplicities in hadronic interactions assuming no ``new
physics''. Comparing that to what we obtained in our estimates for neutrino induced showers
in the string/black hole picture, we can introduce an ``{\em effective atomic number}'':
\BEQ
A_{\rm eff} = \frac{\mbox{\rm multiplicity in the string picture}}{\mbox{\rm extrapolated multiplicity}}
\EEQ 

(Of course, this  estimate of $A_{\rm  eff}$ assumes a simple superposition picture of 
air showers induced by nuclei. In any case, the superposition picture is 
qualitatively correct.)
The result is displayed in the next Table.
\begin{table}[h]
\begin{center}
\begin{tabular}{|c|c|c|c|}\hline
n & 6 & 7 & 8\\ \hline
$A_{\mbox{eff}}$  & 69  & 16 & 4 \\ \hline
\end{tabular}
\end{center}
\caption{Comparison of  proton and black hole induced showers at $E_{L}=3\times 10^{11}$GeV
in terms of an ``effective atomic number''.}
\label{xmaxdiff}
\end{table}

\end{itemize}
It is noteworthy that A.A.~Watson has emphasized over the years that the mass composition
of the extreme energy cosmic rays is uncertain and that, in fact, the data may favor a heavy
composition, see \cite{watsonhere} and references quoted in that paper.
\section{Discussion: Can we tell the difference?}
We presented a scheme of understanding the events induced by extreme energy (trans-GZK)
cosmic rays, based on a scenario involving ``new physics'' as abstracted  from a class
of string models. It is to be emphasized that the scheme presented here is more general 
than any
specific string model known today. It is essential that one rely upon the paradigm of
strongly coupled
string models: the features discussed here, as far as one knows, can be realized in strongly
coupled string models (and their future generalizations to a yet better theory) {\em only}.

It is pleasing that one can can discern some essential differences between the 
string/black hole scheme advocated here and the heavy nucleus scheme as advocated by
Alan Watson, {\em loc. cit.} Both schemes favor ``heavy nucleus''  primaries of the trans-GZK
cosmic ray interactions, see our Table of effective atomic numbers.

However, if the primaries are, indeed, nuclei, they should undergo photo disintegration
in sunlight: this is the celebrated {\em Zatsepin effect}~\cite{zatsepineffect}.
(Briefly, heavy nuclei entering the neighborhood of our solar system interact with
sunlight and undergo  photo disintegration. In the process, one or a few nucleons 
are broken off and get separated from the incoming nucleus. Thus, one can observe
practically simultaneous air showers at detectors separated by distances of
the order of $10^{3}$km or so.) Clearly, no Zatsepin effect 
can be observed  if the primary of a trans-GZK air shower is
a strongly interacting neutrino. Thus, the  presence or absence of the Zatsepin
effect will  differentiate between the (conservative) heavy nucleus
and the (radical?) strongly interacting neutrino schemes, once
 new detectors, such as both sites of the Pierre Auger observatory, will become
operational.

{\bf Acknowledgment.} We thank the organizers of XIII ISVHECRI and Leonidas Resvanis in 
particular, for organizing a very stimulating meeting ``{\em on the sandy beaches of Pylos}'' and for financial support.
It was a pleasure to see old and new friends and to discuss ideas, which will,
probably, lead to the discovery of particle physics going beyond the currently
accepted Standard Model of particle interactions.


\begin{thebibliography}{99}
\bibitem{elbertsommers} J.W.~Elbert and P.~Sommers, Ap.J.~{\bf 441}, 151 (1995).
\bibitem{protheroe} T. Stanev, R.~Engel, A.~M\"{u}cke and R.J.~Protheroe,
Phys.~Rev. {\bf D62}, 093005 (2000).
\bibitem{susan} S.~Kovesi-Domokos, these Proceedings.
\bibitem{strings} G.~Domokos and S.~Kovesi-Domokos, hep-ph/0307099
\bibitem{berezin} V.S.~Berezinsky and G.T.~Zatsepin, Phys. 
Lett. {\bf 28B}, 423 (1969).
\bibitem{prl} G. Domokos and S. Kovesi-Domokos, Phys. Rev. Lett. {\bf 82}, 1366 (1999).
\bibitem{lowscale} E.~Witten,
Nuc.~Phys. {\bf B471}, 135 (1996).
J.~Lykken, Phys.Rev. {\bf D54}, 3693 (1996).  P.~Horava and E.~Witten, Nuc.~Phys. {\bf B475}, 94 (1996)
N.~Arkani-Hamed, S.~Dimopoulos and G.R.~Dvali, Phys.Lett. {\bf B429}, 263 (1998).
I.~Antoniadis, N.~Arkani-Hamed, S.~Dimopoulos and G.R.~Dvali, Phys. Lett. {\bf B4336}, 257 (1998).
\bibitem{blackholes} J.L. Feng and  A.D. Shapere, Phys. Rev. Lett. {\bf 88} 021303 (2002).
\bibitem{amati} D.~Amati, in Proceedings of The Abdus Salam Memorial Meeting, edited by
J.~Ellis, F.~Hussain, T.~Kibble, G.~Thompson and M.~Virasoro. World Scientific Publishing Co.
Singapore (1997).
\bibitem{burgettetal} W.S. Burgett, G.~Domokos and S.~Kovesi-Domokos, hep-ph/0209162
(Expanded version of a paper presented at  ICHEP 2002, Amsterdam, The Netherlands.
\bibitem{coriano} A.~Cafarella, C.~Coriano and T.N.~Tomaras, these Proceedings.
\bibitem{watsonhere} A.A.~Watson, these Proceedings.
\bibitem{zatsepineffect} G.T.~Zatsepin, Dokl. Akad. Nauk (USSR), {\bf 80}, 577 (1951)
N.M. Gerasimova and G.T. Zatsepin, Soviet Phys. (JETP) {\bf 11}, 899 (1960).
G.A.~Medina-Tanko and A.A.~Watson, Astropart. Phys. {\bf 10}, 157 (1999).
\end{thebibliography}
\end{document}